\documentclass[]{aastex631}
\usepackage{amsmath}
\definecolor{DarkGreen}{rgb}{0.0,0.4,0.0}  
\newcommand{\B}[1]{{\color{blue}{#1}}}
\newcommand{\R}[1]{{\color{red}{#1}}}
\newcommand{\G}[1]{{\color{DarkGreen}{#1}}}
\newcommand{\M}[1]{{\color{magenta}{#1}}}
\newcommand{\C}[1]{{\color{cyan}{#1}}}
\usepackage[normalem]{ulem}
\usepackage{soul}
\usepackage{cancel}

\newcommand{\speed}[1]{#1 km\,s$^{-1}$}

\shorttitle{Falling Filament Threads}
\shortauthors{Wu $\&$ Liu}

\graphicspath{{./}{figures/}}

\begin{document}

\title{Falling Threads During Solar Filament Eruptions}

\author[0009-0009-1212-440X]{Yidian Wu}
\affiliation{CAS Key Laboratory of Geospace Environment, Department of Geophysics and Planetary Sciences, \\
University of Science and Technology of China, Hefei, 230026, China}

\author[0000-0003-4618-4979]{Rui Liu}
\affiliation{CAS Key Laboratory of Geospace Environment, Department of Geophysics and Planetary Sciences, \\
University of Science and Technology of China, Hefei, 230026, China}
\affiliation{CAS Center for Excellence in Comparative Planetology\\
University of Science and Technology of China, Hefei, 230026, China}
\affiliation{Mengcheng National Geophysical Observatory\\
University of Science and Technology of China, Hefei, 230026, China}

\author[0000-0002-2559-1295]{Runbin Luo}
\affiliation{CAS Key Laboratory of Geospace Environment, Department of Geophysics and Planetary Sciences, \\
University of Science and Technology of China, Hefei, 230026, China}

\author[0000-0002-9865-5245]{Wensi Wang}
\affiliation{CAS Key Laboratory of Geospace Environment, Department of Geophysics and Planetary Sciences, \\
University of Science and Technology of China, Hefei, 230026, China}
\affiliation{CAS Center for Excellence in Comparative Planetology\\
University of Science and Technology of China, Hefei, 230026, China}

\correspondingauthor{Rui Liu}
\email{rliu@ustc.edu.cn}

\begin{abstract}

Mass drainage is frequently observed in solar filaments. During filament eruptions, falling material most likely flows along magnetic field lines, which may provide important clues for the magnetic structures of filaments. Here we study three filament eruptions exhibiting significant mass draining, often manifested as falling threads at a constant speed ranging between 30--300 km~s$^{-1}$. We found that most of the falling material lands onto the hooked segments of flare ribbons, only a small fraction lands inside the hooks, and almost none lands onto the straight segments of ribbons. Based on these observations we surmise that before eruptions most of the filament mass is entrained by field lines threading the quasi-separatrix layers (QSLs), which wrap around the filament field and whose footpoints are mapped by the hooked ribbons, and that the magnetic reconnection involving these field lines is the major cause of the mass drainage during eruptions. Additionally, the light curves of the hooked ribbons suggest that during eruptions the earlier (later) QSL boundary of filaments is threaded by mass-loaded (depleted) field lines. By assuming that the constant-speed motion is due to a drag force balancing the gravity, we proposed a simplified analytical model to estimate the density contrast of the falling material. The estimated density contrast is then fed into a numerical model, in which filament threads fall along vertical magnetic field lines through a gravitationally stratified atmosphere. The resultant falling speeds, however, are short of observed values, which calls for further investigations.

\end{abstract}


\section{Introduction} \label{sec:intro} 

Solar filaments are cool, dense clouds suspended in the corona. In strong spectral lines such as H$\alpha$, they appear in absorption when viewed against the solar disk, but in emission above the limb, therefore being traditionally termed ``prominences''. This different appearance is a synergy of the opacity and emissivity of the filament itself and the background plasma \citep{Labrosse2010}. Filaments are often categorized by their locations, namely, active-region filaments, intermediate filaments (at the border of active regions), and quiescent filaments (on the quiet sun). Typically, a filament consists of a spine running along the polarity inversion line (PIL), with barbs sticking out from the spine and two ends extending to the lower atmosphere \cite[]{Mackay2010}. Both the spine and barbs consists of threads with widths as thin as 200 km \cite[]{Lin2005}. When observed above the limb, the threads remain horizontal for active-region or intermediate prominences \cite[e.g.,][]{Okamoto2007}, but appear to be vertically oriented in so-called ``hedgerow'' (quiescent) prominences \cite[]{Tandberg-Hanssen1995}, which could be conditioned by the magnetic Rayleigh–Taylor instability \cite[]{Jenkins&Keppens2022}. 

It has been under active investigation as to how the cool, dense filament material is suspended in the hot, tenuous corona. It is generally accepted that magnetic dips, where magnetic field lines are concave upward, may play an important role in supporting filament material against gravity. Dips can be present at the top of magnetic field lines in a sheared arcade \citep{Kippenhahn&Schluter1957} or at the bottom of a magnetic flux rope \citep{Kuperus&Raadu1974}. The latter topology is generally considered to conform with the observations of prominences at the bottom of dark cavities in helmet streamers \cite[]{Low&Hundhausen1995}. Often a U-shaped `horn' in EUV extends upward from the top of the prominence into the cavity \citep{Berger2012}. Further, longitudinal oscillations in filament threads, including the small-amplitude counterstreamings, argue for the presence of magnetic dips \cite[e.g.,][]{Jing2006,Luna&Karpen2012,Luna2014,Awasthi2019}. For partially ionized prominence plasma, a frictional coupling between the neutral and ionized components may provide additional support along with cross-field draining of neutral particles from the prominence \cite[]{Gilbert2002}. Magnetic dips, however, may not be necessary for either the formation or the suspension of prominence material as long as the material is in constant motions, driven by thermal nonequilibrium \citep{Karpen2001}. In addition, \citet{vanBallegooijen&Cranmer2010} proposed that for hedgerow prominences the dense plasma can be supported by the magnetic pressure of a tangled field within the vertical current sheet underneath the flux rope. Numerical modeling and simulation have demonstrated that the distribution of magnetic dips generally agree with that of mass in filaments \cite[e.g.,][]{Aulanier1999,Aulanier&Schmieder2002,Aulanier2002,vanBallegooijen2004,Xia2014,Kang2023}. In the MHD simulation of a prominence-cavity system, \cite{Fan&Liu2019} found that the prominence horns are threaded by twisted field lines containing shallow dips, while the central cavity enclosed by the horns contains twisted field lines without dips. This is different from the simulation by \cite{Xia2014}, where the magnetic structure changes smoothly from the horns to the central cavity. In reality, magnetic configurations of filaments are complicated: a mixture of sheared and twisted fields could be present in the same filament \citep{Aulanier2002,Guo2010}; a combination of sheared arcade and flux rope may co-exist in `double-decker' filaments \cite[]{Liu2012,Awasthi2019}. 

In quiescent prominences, ubiquitous downflows are observed in the form of plasma packets (also termed `knots') descending along vertical threads at a speed of tens of kilometers per second in the plane of sky \cite[]{Engvold1976,Engvold1981,Berger2008,Berger2010,Chae2010,LiuW2012,Li2018,Bi2020}. The acceleration is not constant in time but smaller than that of free fall, ranging from 10 to 200 m~s$^{-2}$. These downflows have been interpreted in different ways. \cite{vanBallegooijen&Cranmer2010} suggested that the prominence plasma flows under the frozen-in condition in a tangled field with a strong vertical component. Within the framework of the Kippenhahn–Schl\"{u}ter solution \cite[]{Kippenhahn&Schluter1957},  \cite{Haerendel&Berger2011ApJ} made an analogy between the falling plasma packets and the droplets of a waterfall in string-like flows with velocities well below free fall due to an aerodynamic friction; they argued that a knot's moving through the horizontal prominence field excites Alfv\'{e}n waves, which produces a drag force and results in the nearly constant-speed downflow. Alternatively, \cite{Chae2010} suggested that a combination of local flux diffusion and ideal MHD relaxation allows plasma to move long distance across horizontal magnetic fields, while \cite{Low2012} invoked a resistive flow across the supporting horizontal field due to a recurrent spontaneous breakdown of the frozen-in condition. \cite{Oliver2014} investigated the dynamics of a dense blob falling in a fully ionized plasma without the effect of magnetic fields. They found that the pressure gradient that acts on the blob to oppose gravity grows with time. Consequently, the blob is initially accelerated but eventually achieves a roughly constant velocity, which is consistent with the observed time-distance diagrams of coronal rain blobs or prominence knots.

Assuming that filament plasma disturbed by eruptive activities primarily flows along inclined field lines, rather than diffuses across horizontal field lines as in quiescent prominences, one may obtain important clues on the filament's magnetic configuration \citep[e.g.,][]{Su&vanBallegooijen2013,Awasthi2019,Awasthi&Liu2019} or how the filament interacts with the ambient field \citep[e.g.,][]{Liu2018}. In particular, counterstreaming flows \cite[e.g.,][]{Luna2012apjl,Alexander2013,ZhouY2020} or longitudinal oscillations \cite[e.g.,][]{Luna2014,Awasthi2019} along the filament spine are consistent with the sheared arcade model, but a bullseye pattern of line-of-sight flows within the cavity \citep[][]{Bak-Steslicka2016} argues strongly for the flux rope model. An eruptive filament's chirality can be inferred from how the draining sites are skewed with respect to the PIL \cite[]{Ouyang2017}. The relation between the filament chirality and the bearing of filament barbs may help discriminate the sheared arcade from the flux rope configuration \cite[]{Chen2014,Ouyang2017}. Magnetic twist of a filament may also be manifested in writhing and/or unwinding motions \citep[e.g.,][]{Alexander2006,Xue2016}. \cite{Awasthi2019} found simultaneous rotational motion about, and longitudinal oscillations along, the filament spine, and suggested that the apparently single filament may in fact possess a double-decker structure comprising a flux rope atop a sheared-arcade. \citet{Awasthi&Liu2019} reported a counter-clockwise rotation of mass motions inside a prominence bubble, with blue-shifted material flowing upward and red-shifted material flowing downward, which may indicate the presence of a kinked flux rope in the bubble emerging from below the prominence.

Filament eruptions are often associated with flares and coronal mass ejections (CMEs). 
In the standard or CSHKP model \citep{Carmichael1964, Sturrock1966, Hirayama1974, Kopp&Pneuman1976} for the classical, two-ribbon flares, initially a filament embedded in a magnetic flux rope starts to rise slowly and stretch the overlying magnetic field, whose legs consequently meet beneath the filament to form a vertical current sheet. Magnetic reconnections at the current sheet produce flare loops and simultaneously add layers of magnetic flux to the rope, which facilitates the acceleration of the rope; the fast-rising rope further stretches the overlying field, which, in return, enhances the reconnection rate. The footpoints of the flare loops constitute the two ribbons located at two sides of the PIL in the lower atmosphere. To explain the hooked morphology at two opposite far ends of the conjugated flare ribbons, the standard model is extended to 3D \citep{2014ApJ...788...60J, 2017JPlPh..83a5301J}. In this model, the two `feet' of the flux rope are outlined by the hooked segments of flare ribbons, which correspond to the footprints of the quasi-separatrix layers (QSLs) that wrap around the flux rope, while the straight segments of flare ribbons correspond to the footprints of the hyperbolic flux tube underneath the flux rope.

Mass draining or settling back to the surface is also frequently observed in filament eruptions, especially in partial or failed eruptions \cite[e.g.,][]{Gilbert2007,Tripathi2009}. Falling plasmas following a prominence eruption generally exhibit larger speeds, broader temperature ranges, and more complex spatial structures than those observed in quiescent prominences \cite[e.g.,][]{Xue2014,Sun2024}. During the filament eruption on 2011 June 7, large amounts of filament fragments fell back to the surface resulting in intense, compact brightening \cite[]{2013Sci...341..251R}. The dissipation of the downflows' kinetic energy is suggested to be the dominant mechanism of the observed surface brightening \cite[]{Gilbert2013}. During the pre-eruption, slow-rise phase of an active-region filament, the ballistic trajectories of downflows toward one end of the filament are well matched by extrapolated NLFFF lines, which suggests that the downflows are oriented along the magnetic field \cite[]{Li2017}. Since downflows during filament eruptions trace the mass-loaded field lines, they may provide clues for the magnetic field of the filaments as well as the eruptive processes. In the sections that follows, we study in detail the properties and behaviors of the falling filament material in three filament eruptions in \S\ref{sec:obs} and discussed the implications of the observations in \S\ref{sec:discussion}.

\section{observation and analysis} \label{sec:obs}
\begin{deluxetable*}{ccccccl}
\tablenum{1}
\tablecaption{Overview of the events\label{tab:events}}
\tablewidth{0pt}
\tablehead{
\colhead{Date} & \colhead{Filament Category} & \colhead{Location} & \colhead{NOAA} & \colhead{Onset} &
\colhead{Flare} & \colhead{CME}
}
\startdata
2011-11-09 & active-region filament & N28E35 & 11343 & 12:50 & M1.1 & Yes \\
2012-02-10 & quiescent filament & N35E15 & - & 18:30 & - & Yes \\
2023-09-14 & intermediate filament & N20W55 & 13425 - 13423 & 06:30 & - &  Yes \\
\enddata
\end{deluxetable*}

We selected three representative events, including the eruptions of one active-region filament, one quiescent filament, and one intermediate filament. The filament eruption in each event exhibits filament material splitting from the rising filament and falling back to the opposite far ends of a pair of conjugate flare ribbons. Basic information of the three events is listed in Table \ref{tab:events}. 

\begin{figure}
\plotone{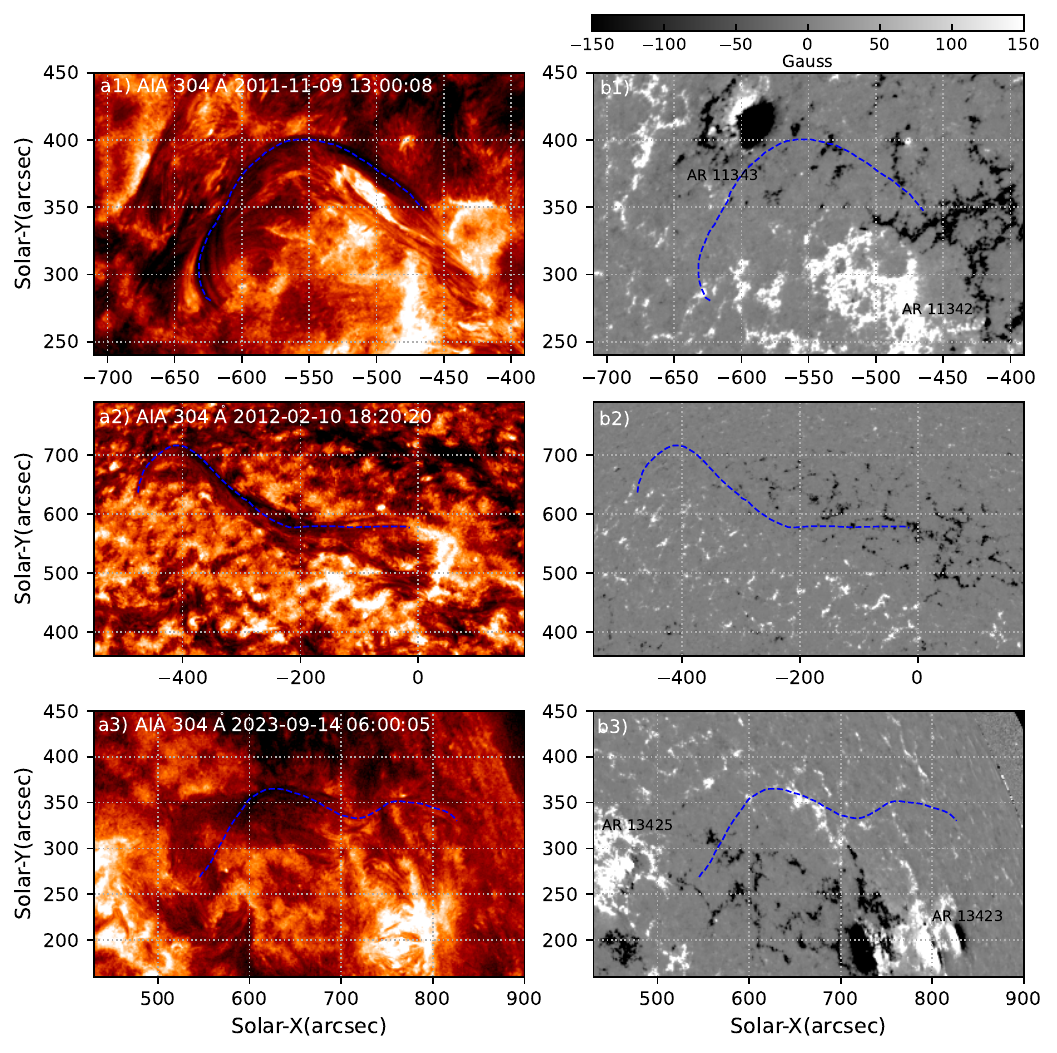}
\caption{Overview of the three events. AIA 304~{\AA} pre-eruption images and HMI line-of-sight magnetograms are shown in the left and right columns, respectively. Filaments in 304~{\AA} are delineated by  blue dashed lines, which are superimposed in magnetograms. 
\label{filaments_mag}}  
\end{figure}

These events are observed in multi-wavelengths by the Atmospheric Imaging Assembly \cite[AIA;][]{2012SoPh..275...17L} onboard the Solar Dynamics Observatory \cite[SDO;][]{2012SoPh..275....3P} at a $0\farcs6$ pixel scale and a temporal cadence of 12~s for the seven EUV bands and a cadence of 24~s for the two UV 1600 and 1700 \AA\ passbands. We also use images taken by the Extreme Ultraviolet Imager \cite[EUVI;][]{2008SSRv..136...67H} onboard the ``Ahead'' and ``Behind'' satellites of the Solar Terrestrial Relations Observatory \cite[STEREO;][]{2008SSRv..136....5K}. By pairing EUV images taken at different viewing angles, it is possible to reconstruct the 3D trajectories of the falling material \cite[e.g.,][]{Gilbert2013}. In Figure~\ref{filaments_mag}, the three filaments as observed in the AIA 304~{\AA} passband are shown side by side with the corresponding line-of-sight magnetograms obtained by the Helioseismic and Magnetic Imager \cite[HMI;][]{Scherrer2012} onboard SDO.

\subsection{Landing sites of filament material} \label{subsec:pos}

We focus on the landing sites of falling filament material in relation to the flare ribbons. The falling material is best observed in the running difference images of AIA 335 or 171~{\AA} passbands, while the flare ribbons are visible in AIA's UV passbands for the eruption of the active-region filament, but the eruptions of the quiescent and intermediate filaments are only visible in the EUV passbands (e.g., 304~{\AA}). Below we briefly describe each individual event.

\subsubsection{2011-11-09 event}

The filament in the 2011-11-09 event is rooted in the NOAA active region (AR) 11342 (Figure~\ref{filaments_mag}). But the M1.1-class flare associated with the filament eruption is registered to occur in AR 11343 in the GOES flare catalog. This is because AR 11342 is closely adjacent to AR 11343 in the northeast, and the flare ribbon located in the negative-polarity region extends all the way into the major sunspot of AR 11343 (see the animation accompanying Figure~\ref{fig:11}), as observed in the AIA 1600~{\AA} passband. The relation between the falling filament material and the flare ribbon is well observed on the hooked ribbon in the east. We identify the falling filament material in running difference images of the AIA 335~{\AA} passband, and mark the ends of both bright or dark falling threads with cyan circles as the landing sites. By comparing the falling filament material in 335~{\AA} and the flare ribbons outlined by red lines from 1600~{\AA} images (Fig.~\ref{fig:11}), one can see that most of the filament plasma landed right onto the hooked ribbon. 

As soon as the filament starts to rise at about 13:05 UT, two flare ribbons begin to develop in 1600~{\AA}; meanwhile, falling filament material, which is subject to some sort of heating, is manifested as a mixing of brightening and dark threads in 335~{\AA} (Figure \ref{fig:11}(a1), (b1), and the accompanying animation). Swirling and rolling motions of the threads about the filament leg are observed simultaneously with downflows along the threads that end at the hooked section of flare ribbons. Although the hooked region shrinks as the hooked ribbon undergoes a westward sweeping motion, the falling material keeps landing onto the front of the hooked ribbon. Starting from about 13:20 UT, some material is detached from the main body of the filament and falls along curved paths toward an area to the east of the flare ribbons, resulting in a side ribbon in 1600~{\AA}, almost attached to the hooked ribbon in the east (Figure \ref{fig:11}(a2)). Later when the filament rises to higher altitudes, the falling filament material is mainly manifested as dark threads falling along the legs of the filament (Figure \ref{fig:11}(a3)), the vertical sections that are still attached to the surface during the eruption. Meanwhile, a few threads may land inside the hook and the impact of the falling material causes a few discrete brightening points. On the other hand, we don't observe any significant draining landing on the straight segments of the flare ribbons, at which the post-flare loops are anchored.

\subsubsection{2012-02-10 event}

The 2012-02-10 event occurs in a quiet region. Filament material is observed to flow along both legs back to the surface during the eruption. The flare ribbons are visible in AIA 304~{\AA} but not in 1600~{\AA}. As the ribbons are not as well defined as those in active regions, we employed coronal dimmings, as a proxy of the footpoints of the eruptive structure, to facilitate the identification of the hooked ribbons enclosing the footpoints \cite[e.g.,][]{Wang2017,Gou2023,Wang2023}. Specifically, we identify a pair of coronal dimming regions (blue contours in Fig.~\ref{fig:12}) by extracting pixels dimmer than sixty percent of the background intensity (averaged over half an hour before the eruption) during the eruption. We then outline the flare ribbons in 304~{\AA} by red lines and overlay them on running difference images of the 171~{\AA} channel. The landing sites of the falling material are again marked with cyan circles. Zooming in on both hooked regions, one can see that the dimming regions, which map the footpoints of the eruptive structure, are indeed enclosed by hooked ribbons. And with the aid of the animation accompanying Fig.~\ref{fig:12}), one can see that filament material constantly flows downward along threads onto the hooked ribbon even though the hook morphology is changing with time.  

\subsubsection{2023-09-14 event}
In the 2023-09-14 event, as the filament rises, strands of filament material exhibit twisting and rolling motions, while the majority of falling material moves downward along the eastern leg of the filament and lands on the southern hooked ribbon. Similar to the 2011-11-09 event, the hooked region shrinks as the hooked ribbon undergoes a westward sweeping motion, but falling material keeps landing on the leading front of the hooked ribbon (see the accompanying animation of Figure \ref{fig:13}). A few falling threads land inside the hooked region, and their ends on the surface are brightened (Figure \ref{fig:13}(a3)), as a result of density or temperature increase due to the impact. Again we do not notice any significant mass draining that is directed toward the straight segments of flare ribbons, which correspond to the footpoints of post-flare loops as observed, e.g., in 171~{\AA}.

\subsection{Dynamics of falling\label{subsec:speed}}
\begin{figure}
\plotone{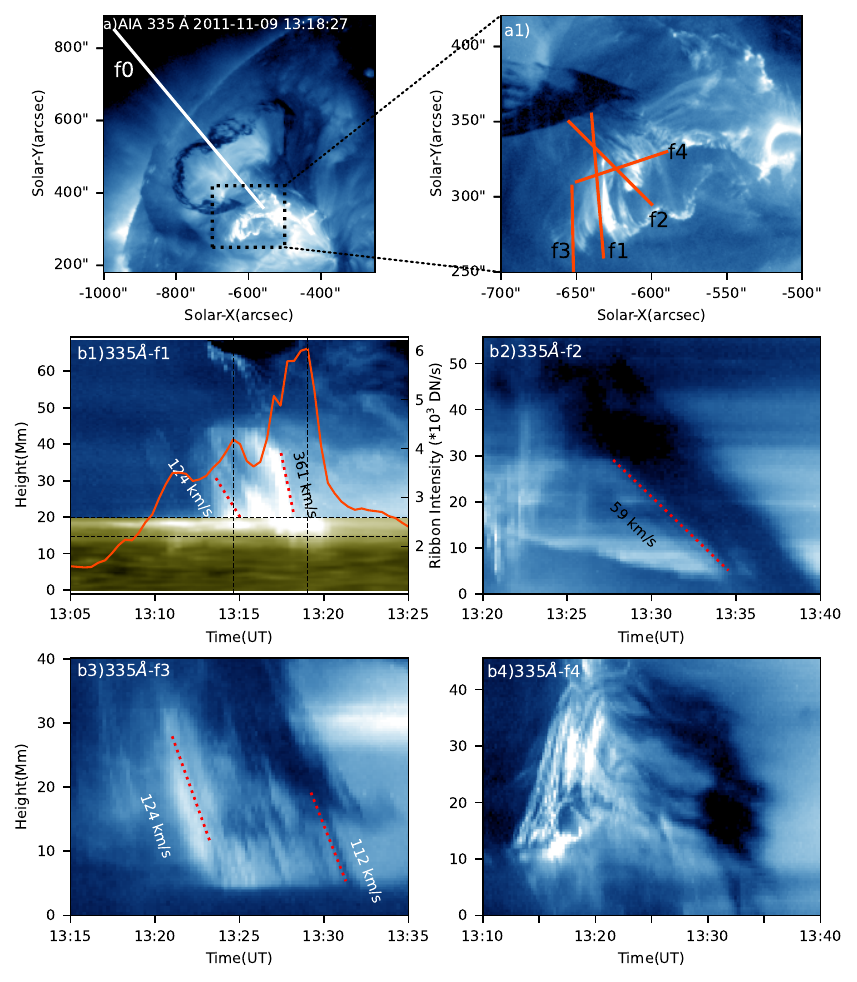}
\caption{Estimation of falling speeds with stack plots derived from AIA 335 \AA\ images for the 2011-11-09 event. Panel (a) shows a 335 \AA\ image taken at 13:18 UT; (a1) zooms into the box region in (a) to show the flare ribbon at the eastern leg of the filament. The virtual slit f0 (solid white line in (a)) is oriented along the direction of the filament eruption; the slits f1--f3 (solid orange lines in (a1)) are oriented along the falling directions of the filament material; f4 (solid orange line in (a1)) is oriented along the east-to-west swirling direction of the filament material. In the corresponding time-height stack plots, dashed lines mark the episodes of mass drainage, whose speeds are estimated by linear fitting. Particularly, (b1) combines two stack plots from the same slit f1 in 335 and 1600~{\AA}. The light curve of the ribbon segment intersected by the slit in 1600~{\AA}, in between the two horizontal dashed lines, is given by the orange curve. The peaks of the light curve associated with falling threads are marked by vertical dashed lines. 
\label{fig:21}} 
\end{figure}

\begin{figure}
\plotone{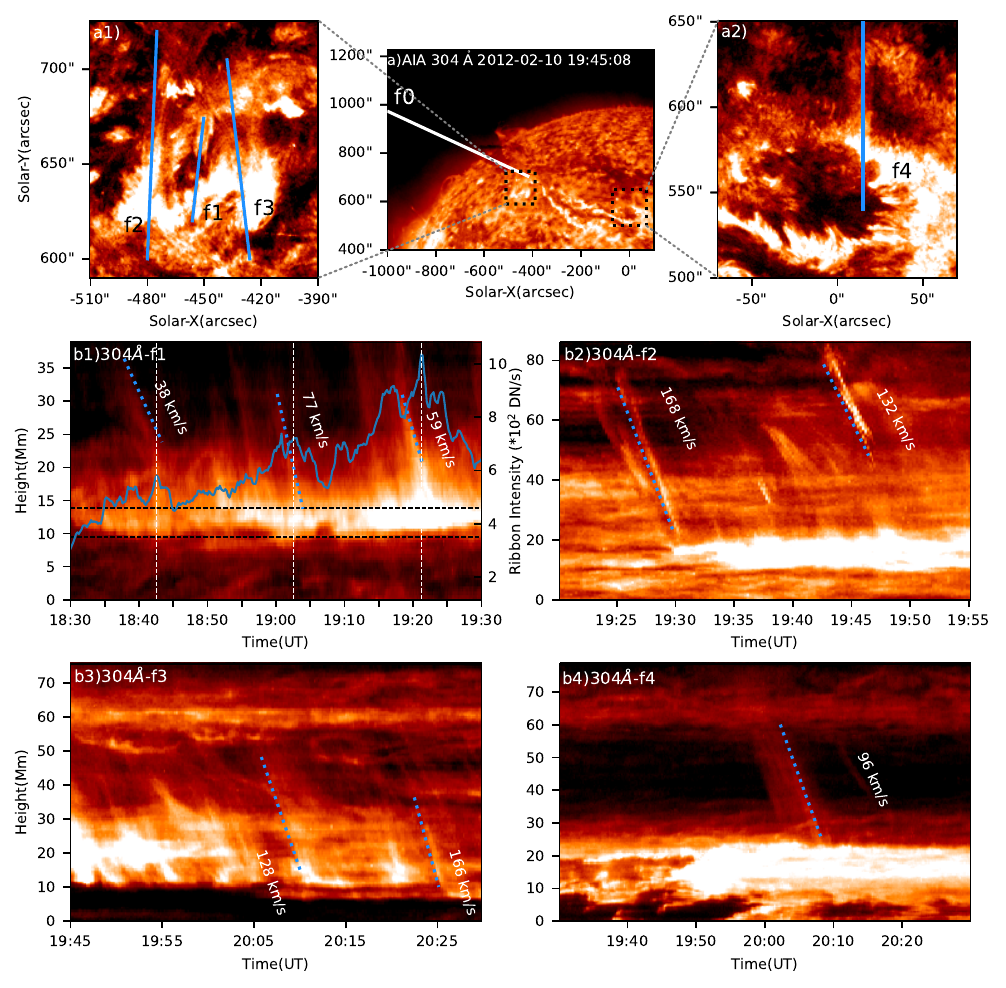}
\caption{Estimation of falling speeds with stack plots derived from AIA 304 \AA\ images for the 2012-02-10 event. Similar to Figure \ref{fig:21}, the slit f0 is oriented along the direction of the filament eruption; the slits f1--f4 are oriented along the falling directions of the filament material. In (b1) the stack plot is superimposed by the light curve (blue) of the ribbon segment intersected by the slit in 304~{\AA}, in between the two horizontal dashed lines. The peaks of the light curve associated with falling threads are marked by vertical dashed lines. 
\label{fig:22}} 
\end{figure}

\begin{figure}
\plotone{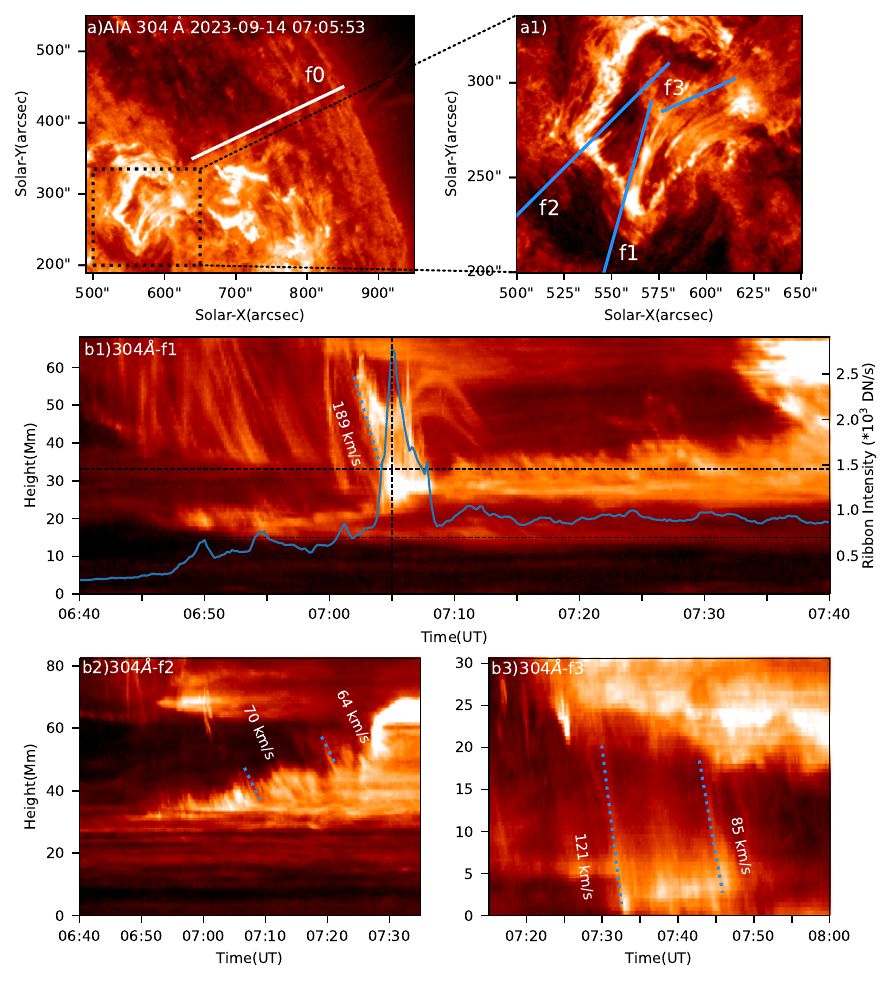}
\caption{Estimation of falling speeds with stack plots derived from AIA 304 \AA\ images for the 2023-09-14 event. Similar to Figure \ref{fig:22}, the slit f0 is oriented along the direction of the filament eruption; the slits f1--f3 are oriented along the falling directions of the filament material. In (b1) the stack plot is superimposed by the light curve of the ribbon segment intersected by the slit, similar to Fig.~\ref{fig:22}.
\label{fig:23}} 
\end{figure}

We apply a series of virtual slits to study the kinematics of the falling material. These include the slits approximately parallel to the projected falling directions, a slit perpendicular to the falling direction for the 2011-11-09 event (`f4'), and a slit to track the rising filament (`f0') for each event (Figs.~\ref{fig:21}-\ref{fig:23}). By arranging the slices taken from the slits in a chronological order, we make a stack plot for each slit, which gives the time-height evolution of the features tracked by the corresponding slit. 

\begin{figure}
\plotone{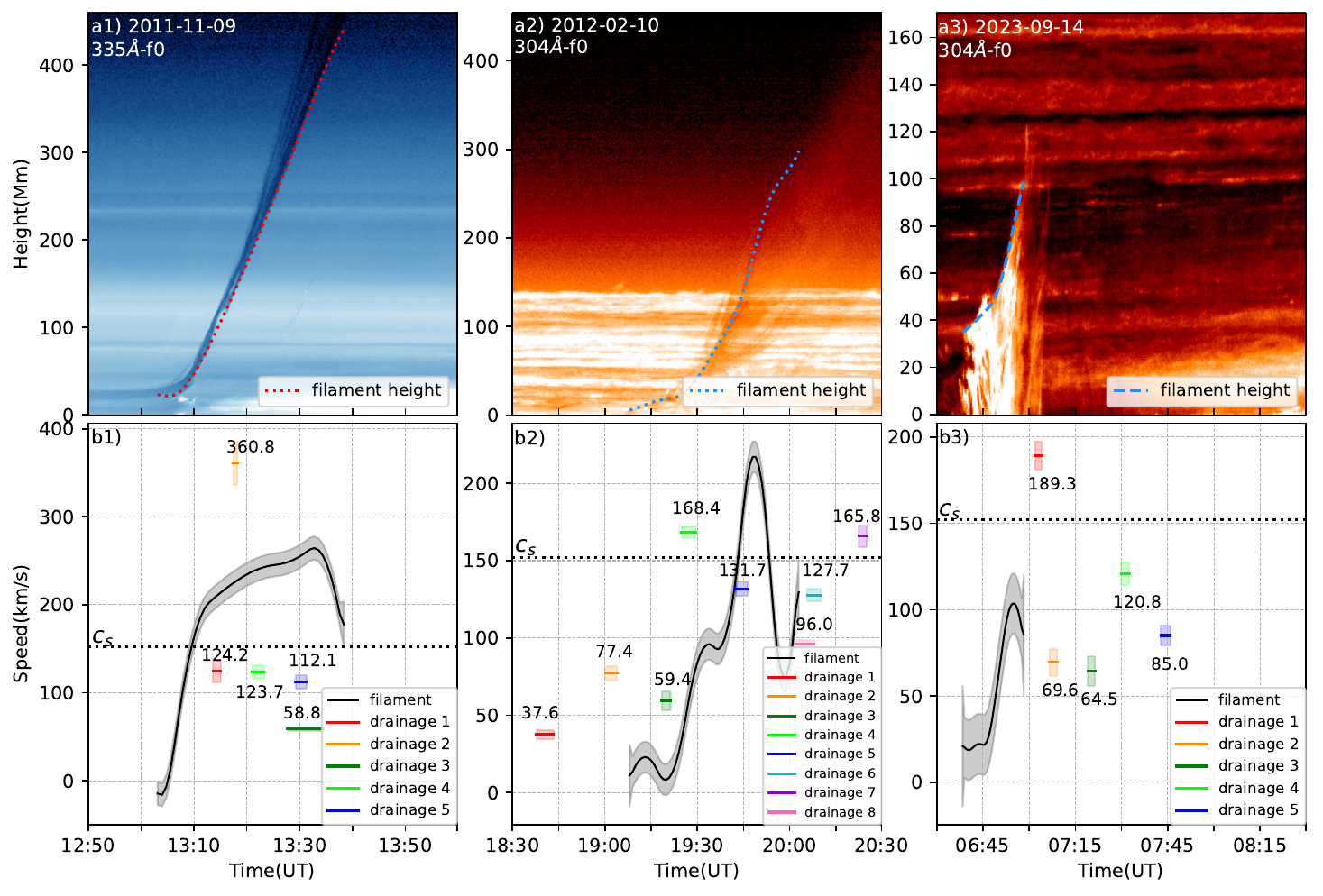}
\caption{Speeds and timing of falling threads in relation to the filament eruptions of interest. The top panels show the trajectories of the eruptive filaments obtained from the virtual slit f0 in Figs.~\ref{fig:21}-\ref{fig:23}. The filament speeds are calculated by the time derivative of the height-time profiles. The falling speeds are obtained by linearly fitting the streaks of the falling threads in the stack plots as illustrated in Figs.~\ref{fig:21}-\ref{fig:23}. The speed uncertainties are indicated by shaded areas. The horizontal dashed line marks the sound speed $c_s\simeq 152$ km s$^{-1}$ in 1~MK plasmas.
\label{fig:2}} 
\end{figure}

Surprisingly, filament material seems to fall down at a constant speed as revealed by bright or dark streaks in the stack plots acquired from the slits parallel to the falling directions, including the falling material landing inside the hooked ribbon in the 2023-09-14 event (Figure \ref{fig:23}(b3)). Along the same slit, some filament material seems to fall at similar projected speeds since the streaks in the same stack plot are inclined with similar slopes.
By linearly fitting the streaks we obtain the falling speeds. We select a few distinctive episodes of mass drainage and summarize the estimated speeds in Figure \ref{fig:2}. The vertical bars indicate the speed uncertainty, while the horizontal bars represents the duration for each episode of falling. The speed uncertainty is calculated by assuming a 2-pixel measurement error in delineating the rising streaks of eruptive filaments or falling streaks of mass drainage.

The projected speeds of the falling material ranges from 30 to 300 kilometers per second. Most of the mass drainage occur during the fast rising phase of the filament eruption with a speed slower than the sound speed of the 1~MK plasma ($c_s = \sqrt{\gamma k_B T/ \mu m_p}$, where $\mu = 0.6$ is the mean molecular weight of fully ionized coronal plasma). Occasionally the estimated falling speed exceeds both the sound speed and the filament's rising speed. The falling speeds are subject to large uncertainties, because generally the falling material does not move exactly along the slit. Sometimes the falling threads undergo a swirling motion with their lower ends apparently slipping along the hooked ribbon (see also the animation accompanying Fig.~\ref{fig:11}), and an example is demonstrated by the stack plot derived from the slit across the threads (Fig.~\ref{fig:21}(b4)).  

\subsubsection{A simplified model on constant-speed falling} \label{subsubsec:model}
Despite the uncertainties, we conclude that overall the falling material moves largely at a constant speed, indicating that a force balance is achieved on the way of falling. This is consistent with \citet{Gilbert2013}, who reconstructed the 3D paths of falling material by performing triangulation measurements with STEREO and SDO observations. We expect that the filament material is still frozen in with the magnetic field that dominates in the corona, through interactions between the ionized atoms and the neutral ones. Hence, the falling material must largely move along magnetic field lines, without sensing any force exerted by the magnetic field, and the gravity must be balanced by the interplay between the falling material and the surrounding plasma, which is considered as a drag force. In hydrodynamics, drag force acts in a direction opposite to the relative motion of an object in a fluid. In the interplanetary medium, it is the drag force acting on the interplanetary coronal mass ejection (ICME) that leads to the ‘equalisation’ of the ICME and solar wind velocities \citep{2004On}. 

Usually the drag force depends on the properties of the fluid and on the size, shape, and speed of an object moving through the fluid. It can be expressed as follows,
\begin{equation}
F_D =  \frac{1}{2} \rho_e v^2 c_d A ,
\label{eq1}
\end{equation}
where $\rho_e$ is the density of the ambient fluid, with subscript $i$ and $e$ indicating \emph{internal} and \emph{external} values for the object, respectively. $v$ is the speed of the object relative to the fluid, $A$ is the object's cross-section area, and $c_d$ is the dimensionless drag coefficient. In aerodynamics, for objects with certain simple geometry, $c_d$ can be calculated analytically \citep{1987flme.book.....L}, but there is no analytic formula for $c_d$ in magnetohydrodynamics (MHD; \citealt{2022JGRA..12728744L}). Simulations for ICMEs have shown that $c_d$ is of order unity and approximately constant between the Sun and 1 AU \citep{2004On}. In this study, we take $c_d$ as unity for simplicity.

\begin{figure}
\plotone{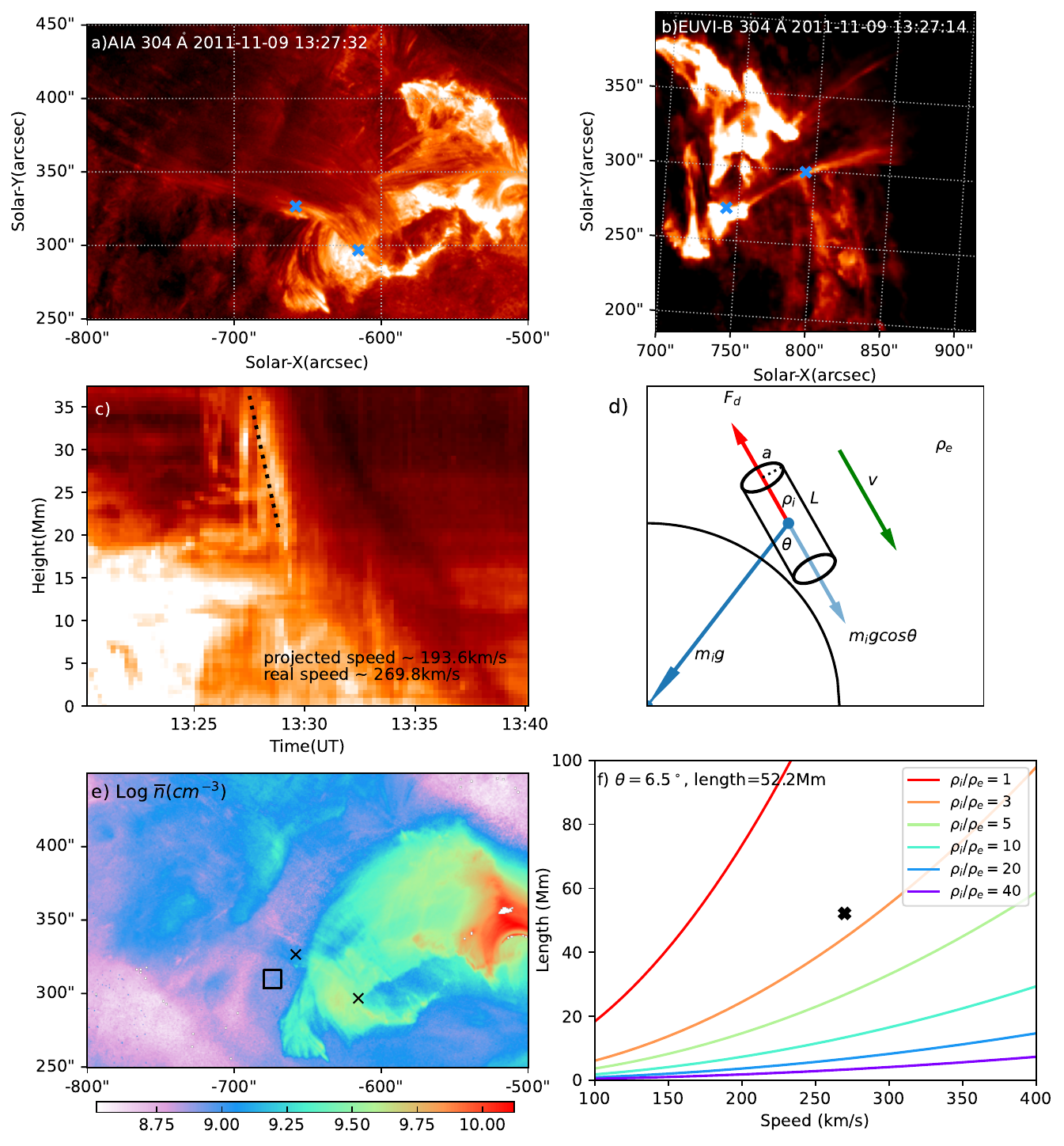}
\caption{Diagnostics on the density contrast of falling filament threads. (a \& b) Paired images of the 2011-11-09 event observed simultaneously by SDO/AIA (a) and STEREO-B/EUVI (b). The blue crosses denote the visually selected two ends of a segment on a falling thread observed by both spacecrafts. (c) Time-height diagram produced with a slit oriented along the falling thread. (d) Schematic diagram of an inclined magnetic flux tube with length $L$ and radius $a$. The dark blue arrow points from the center of the flux tube to the center of the Sun, indicating the direction of the gravity force. The light blue arrow indicates the effective gravity force that is parallel to the magnetic field. The red arrow indicates a drag force that balances the effective gravity force. The motion of the tube is marked by the green arrow. (e) density map derived from a differential emission measure analysis. The average density in the rectangular area is taken as the medium density outside the falling thread. (f) speed-length diagram derived from Eq.~\ref{eq3} with a fixed tilt angle of $6.5\arcdeg$ and different density ratios. The black cross denotes the measured speed and length of the thread under investigation. 
\label{fig:recon}} 
\end{figure}

Based on observations that the falling material takes the appearance of bright or dark threads, we model the threads as magnetic flux tubes characterized by radius $a$ and length $L$ (Fig.~\ref{fig:recon}d), without taking into account the changing cross-sectional area along the length of the tube. These cylindrical flux tubes fall obliquely, with the drag being balanced by the component of gravity parallel to the moving direction:
\begin{equation}
F_D =  m_i g \cos\theta ,
\label{eq2}
\end{equation}
in which $g = 0.274$~km~s$^{-2}$ is the solar gravitational constant and $\theta$ is the tilt angle with respect to the radial direction (Fig.~\ref{fig:recon}d). The tube mass can be written as $m_i = \rho_i AL$, with $\rho_i$ referring to the mass density inside the tube. Inserting this expression and Eq.~(\ref{eq1}) into Eq.~(\ref{eq2}), the tube cross-section area cancels out and we obtain the density contrast:
\begin{equation}
\rho_c  = \frac{\rho_i}{\rho_e} =  \frac{c_d}{2g} \frac{v^2}{L \cos\theta}, \label{eq3}
\end{equation}
where the length $L$, speed $v$, and tilt angle $\theta$ can be estimated from observations. The density contrast is sensitive to the tilt angle in the denominator, especially when the flux tube is nearly horizontal ($\theta\rightarrow90\arcdeg$). Moreover, the heavier is the tube, the faster it falls.

To mitigate the projection effects, we perform triangulation measurements of the falling material, combining STEREO and SDO observations. Utilizing the python routine \texttt {triangulate.py} \cite[]{2023SoPh..298...36N} we reconstruct the 3D orientation and length of the threads. Limited by the low cadence of EUVI images and sometimes the unfavorable separation angle between the two spacecrafts, only a few threads are reconstructed in 3D by the stereoscopic approach. Figure \ref{fig:recon} shows an example from the 2011-11-09 event. The separation angle between the ``Behind'' satellite of STEREO and SDO is $102.9\arcdeg$. The blue crosses mark the two ends of the thread from both perspectives; the length of the thread is taken by the distance between the two marks. With a virtual slit along the thread between the two marks, we obtain a stack plot (Fig.~\ref{fig:recon}c) to estimate the average plane-of-sky speed of the falling material, and then convert it to real speed with 
\[\frac{v_\mathrm{real}}{v_\mathrm{POS}} = \frac{L_\mathrm{real}}{L_\mathrm{POS}}.\] 
We estimated that $L\approx 52.2$ Mm, $v_\mathrm{real} \approx 269.8$ km~s$^{-1}$, $\theta \approx 6.5\arcdeg$, which yields a density contrast $\rho_c\sim$ 2.6. The mean electron number density in the background corona near the thread is estimated through a differential emission measure (DEM) analysis \citep{MCheung2015}, so that
\begin{equation}
n_e = \sqrt{\frac{\int_T \mathrm{DEM}(T)\,dT}{H}} \approx 8 \times 10^8 \ \mathrm{cm}^{-3},
\end{equation}
where $H= \frac{k_B T}{\mu m_p g}$ refers to the hydrostatic scale height, with the proton mass $m_p = 1.67 \times 10^{-24}$~g and mean molecular weight $\mu=0.5$ for the 1~MK hydrogen plasma. Thus, the electron density in the falling flux tube is estimated to be about $n_i\approx 2 \times 10^9$ cm$^{-3}$. These numbers fall in the range given in the literature, i.e., the electron number density of prominence plasma is typically in the order of $10^9$--$10^{11}$ cm$^{-3}$ \citep{Labrosse2010}, and that of coronal plasma is in the order of $10^8$--$10^{9}$ cm$^{-3}$ in the lower corona \cite[]{Fludra1999}. Assuming that most of hydrogen atoms in the flux tube are fully ionized, we obtained the mass density of the falling flux tube as $\rho_i\approx n_i m_p\approx3.3\times10^{-15}$ g\,cm$^{-3}$ for a gas consisting of hydrogen only. However, one must keep in mind that $\rho_i$ is not the density of prominence material, but the average mass density of the flux tube, due to an unknown filling factor, which is estimated to range in [0.001, 0.2] from various observations \citep{Labrosse2010} and is about 0.1--0.15 in the simulation result of \citet{ZhouY2020}.

\subsubsection{Numerical experiments on the falling dynamics}
Here we aim to better understand the dynamics of falling by performing a one-dimensional simulation along the falling path, following \cite{Oliver2014}. The evolution of plasma density ($\rho$), velocity ($\mathbf{v}$), and pressure ($p$) is described by the conservation of mass, momentum, and energy in a fully ionized hydrogen plasma \citep[Eqs.~(1)-(3) in][]{Oliver2014}. Assuming that $(\rho,p,\mathbf{v})$ depend only on $z$ and $t$ and setting positive $z$ opposite to the direction of falling, we have
\begin{equation}
    \begin{split}
        \frac{\partial\rho}{\partial t} = -v\frac{\partial\rho}{\partial z}-\rho\frac{\partial v}{\partial z} \\
        \rho\frac{\partial v}{\partial t}=-\rho v\frac{\partial v}{\partial z}-\frac{\partial p}{\partial z}-\rho g\\
        \frac{\partial p}{\partial t} = -v\frac{\partial p}{\partial z} -\gamma p\frac{\partial v}{\partial z}
    \end{split}
\end{equation}
Magnetic field is taken out of the equations because filament mass is assumed to flow along field lines, and non-ideal effects are omitted for simplicity. For material falling obliquely, the effective gravitational acceleration is corrected by a factor $\cos\theta$, as described before.   

Initially, the static, isothermal atmosphere is gravitationally stratified: 
\begin{equation} \label{eq:background}
p(z, t=0) = p_{0}\,e^{-z/H} ,\quad
\rho(z, t=0)  = \rho_{0}\,e^{-z/H},
\end{equation}
with the pressure and density at the coronal base being specified by the ideal gas law
\begin{equation}
    p_{0}= (k_\mathrm{B}/ \mu m_p) \rho_{0} T_{0}.
\end{equation}
The model atmosphere does not include the density jump at the base of the corona, since this study focuses on the dynamics of falling material in the corona, and the observations do not have sufficient spatio-temporal resolution to reveal how the falling material reacts to the density jump. 

A thread of the length $2\Delta$ and the density distribution
\begin{equation}
    \rho_{th}(z, t=0) = \rho_{th,0}\,\exp \left[-\left( \frac{z-z_0}{\Delta}\right)^2 \right],
\end{equation}
is superimposed to the above static atmosphere at a height $z_0$, with $\rho_{th,0}$ the density at the middle of the thread. Adopting the same density distribution as given by Eq.~14 in \cite{Oliver2014}, we essentially model the thread as a highly stretched blob; the thread length is much larger than the blob length (0.5 Mm) in \cite{Oliver2014}. Now, the initial density is the sum of the background density in Eq.~\ref{eq:background} and the thread density $\rho_{th}(z, t=0)$, but the initial pressure is still given by Eq.~\ref{eq:background}. In a mechanical non-equilibrium the thread starts to fall. 

We use the parameters of the reconstructed thread in \S\ref{subsubsec:model}, i.e., $z_0=37$ Mm, $\rho_{th,0}\approx 3.3\times10^{-15}$ g\,cm$^{-3}$, and $\Delta =26$ Mm, as the initial condition, and set the initial temperature of the atmosphere $T_0=2$ MK, which gives the scale height $H \approx 119$ Mm, and at the coronal base $\rho_{0} \approx 1.8\times10^{-15}$ g\,cm$^{-3}$ and $p_{0} \approx 0.06$ Pa. The simulation domain ranges from -200 Mm to 350 Mm; the grid spacing in the range of -70 $\sim$ 200 Mm is 0.01 Mm, but 0.1 Mm elsewhere. We follow the boundary conditions in \citet{Oliver2014}: the density and pressure are fixed at the top and bottom boundaries, where $\partial v/\partial z=0$. The PDE2D code \citep{sewell2005numerical} with Galerkin’s method is used for the simulation. 

\begin{figure}
\plotone{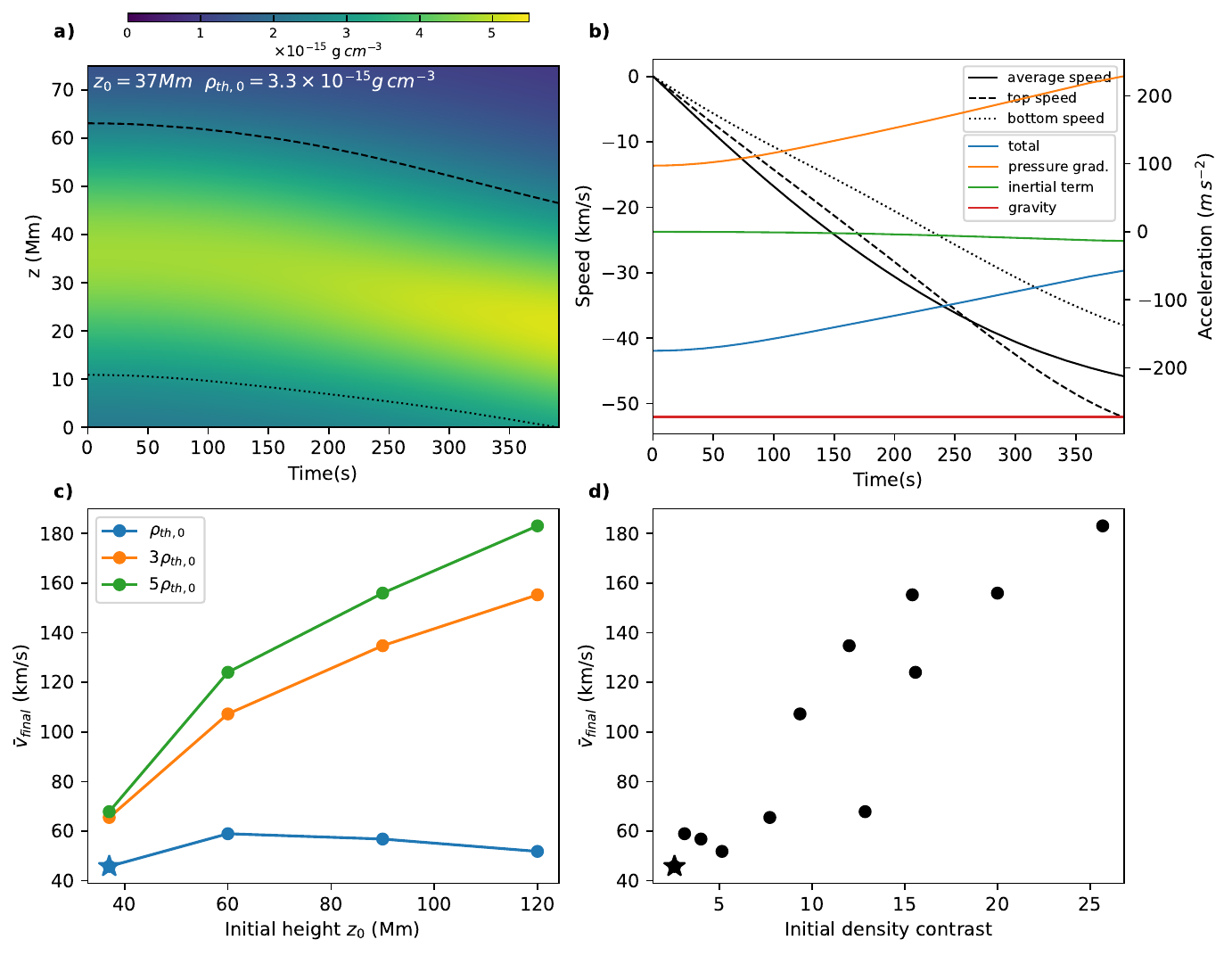}
\caption{Dynamics of the simulated falling threads. The result using the measured thread parameters as input is shown in the top panels, and the measured values are highlighted by star symbols in the bottom panels. In (a) the density distribution of the falling thread is indicated by the color bar at the top. The dashed (dotted) line outlines the trajectory of the top (bottom) of the thread. (b) shows the top, bottom, and average speeds of the falling thread (scaled by the left $y$-axis) and the accelerations of the thread as a whole as caused by different forces acting on it (scaled by the right $y$-axis). (c) shows the final average speed $\Bar{V}_\mathrm{final}$ for threads with different initial central density ($[3.3, 9.9, 16.5]\times10^{-15}$ g\,cm$^{-3}$) and height (37, 60, 90, 120 Mm). (d) reorganize the data points in (c) to show $\Bar{V}_\mathrm{final}$ as a function of the initial density contrast.
\label{fig:simulation}}
\end{figure}

The simulation result are shown in Figure~\ref{fig:simulation}. The top and bottom ends of the falling thread, where the mass density drops to $1/e$ of the densest point in the thread, are outlined by the dashed and dotted lines in Figure~\ref{fig:simulation}a. As time progresses, the thread is slightly shortened and and its density increases probably due to compression by the surroundings. The material falls under the competition between the pulling of the gravity and the drag of pressure gradient, eventually reaching a maximum speed of \speed{45.8} at $z=0$ Mm, which falls far short of the observed speed of \speed{269.8}. The top of the thread falls faster than its bottom (Fig.~\ref{fig:simulation}b), which explains the shortening length. Moreover, unlike the observed constant-speed falling, the thread in simulation is accelerated with 1/5--1/2 of the Sun's gravitational acceleration, which suggests that some ignored effects such as viscosity may contribute to the drag.

We further investigate how the final average speed of a thread, $\Bar{V}_\mathrm{final}$, which is averaged over the thread when its bottom end reaches the coronal base, depends on the initial states including the height and density (Fig.~\ref{fig:simulation}(c \& d)). To take into account the filling factor (\S~\ref{subsubsec:model}), we set the thread central density $\rho_{th,0}$ to 3 to 5 times denser than the observation. The initial density contrast is taken as the ratio of the central density $\rho_{th,0}$ over the ambient density of the model atmosphere. With a series of initial heights ($z_0=[37, 60, 90, 120]$~Mm;  Fig.~\ref{fig:simulation}c), we obtained different initial density contrasts ranging from 3 to 30. Fig.~\ref{fig:simulation}c shows that $\Bar{V}_\mathrm{final}$ increases with both increasing $z_0$ and $\rho_{th,0}$ as expected for the gravity-dominated motions. Fig.~\ref{fig:simulation}d further demonstrates that $\Bar{V}_\mathrm{final}$ is positively correlated with the initial density contrast, which is similar to the conclusion given by \citet{Oliver2014}. These experiments suggest that we may have underestimated the initial density contrasts and heights of the falling threads, as they are better observed in the lower atmosphere, especially against the flaring plasma (see the animations accompanying Figs.~\ref{fig:11}-\ref{fig:13}). If starting to fall at higher altitudes, the threads would also have larger density contrasts against the surroundings. The combination of the above two factors may help the falling threads in the experiments achieve a final speed as fast as the observed ones.

\subsection{Timing of falling \label{subsec:light}}

It is suggested that the impact of mass drainage may heat up the chromospheric plasma by converting the kinetic energy of falling material to thermal energy \citep{Gilbert2013}. By extending the virtual slits along the falling threads to cross over the flare ribbon upon which the falling material lands (Figs.~\ref{fig:21}(b1), \ref{fig:22}(b1), and \ref{fig:23}(b1)), we found that indeed the peaks on the light curve of the ribbon segment intercepted by the slit are closely associated with episodes of mass drainage. However, there is no one-on-one association between mass drainage and ribbon brightening, and the landing of the falling material may take place either before or after the peak of the ribbon brightening. Hence, these discrete episodes of mass drainage is more likely a byproduct of the magnetic reconnection that proceeds in a bursty manner between the mass-carrying flux rope and the ambient field than a direct cause of the ribbon brightening. 

Below we investigate the light curve of flare ribbons to examine whether the falling materials have produced extra brightening in the chromosphere as they land on the flare ribbons as conventionally identified in UV images. The ribbon brightness is examined for the 2011-11-09 and 2023-09-14 events in AIA 1600~{\AA} images, but not the 2012-02-10 event as its ribbons are not visible in 1600~{\AA}. Stacking up 1600~{\AA} images over the entire flaring period, we obtain a `synoptic' map of flare ribbons, in which each pixel is shown by its maximum intensity during the flaring period, so that the ribbon-swept area is highlighted (e.g., Fig.~\ref{fig:31}a). The light curve for each ribbon is given by the average intensity of pixels brighter than the background in the outlined area, and each ribbon is further divided into hooked and straight ribbons. By visually examining the time-lapse animation of AIA images, we further divide the hooked ribbon into one segment associated with significant material drainage and one without. 

In the 2011-11-09 event, the light curves of both the hooked (R1) and straight (R2) ribbon in the south peak at about 13:20 UT (Fig.~\ref{fig:31}b). The conjugated ribbon in the north (R4) lacks a major peak. The two segments of the hooked ribbon in the south, with (R1a) and without (R1b) significant mass drainage, peak at similar intensities, except that R1b shows a second peak (Fig.~\ref{fig:31}c). Different segments of the straight ribbon also peak at different times (Fig.~\ref{fig:31}d). The segment R2a peak at about 13:10 UT, 10 minutes earlier than the segments R2b and R2c. In addition, the light curve of the side ribbon (R3), which is the landing site of mass falling outside the major flare ribbons, shows a peak much weaker and later than those of the major flare ribbons (Fig.~\ref{fig:31}b). We thus argue that the brightening at the side ribbon is primarily caused by the impact of falling filament material, while the brightening at major flare ribbons is mainly caused by magnetic reconnection, since the two segments of the hooked ribbon have similar brightness, regardless of whether there is significant mass drainage upon the ribbon segment.  

\begin{figure}
\plotone{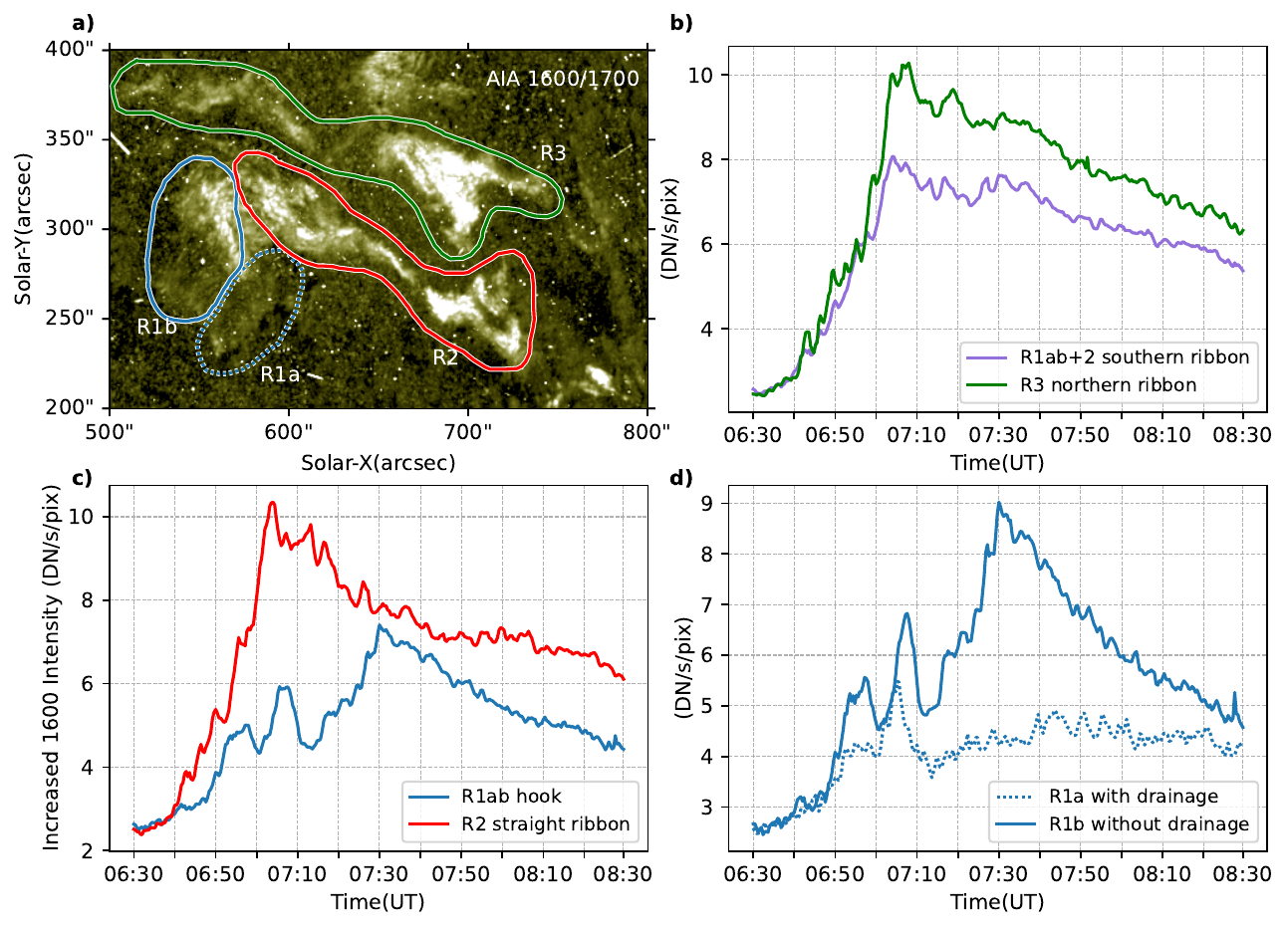}
\caption{Light curves of different segments of the flare ribbons in the 2023-09-14 event. (a) synoptic map of flare ribbons in the ratio of 1600/1700 passbands to remove plages. (b) light curves for southern and northern ribbon, respectively. (c) light curves for the hooked and straight ribbon in the south. (d) light curves for segments on the hooked ribbon with and without drainage.
\label{fig:33}} 
\end{figure}

In the 2023-09-14 event, the light curves of the southern and northern ribbon peak simultaneously at 07:07 UT (Fig.~\ref{fig:33}b). But the light curve of the straight ribbon in the south (R2) is distinct from that of the hooked ribbon (R1; Fig.~\ref{fig:33}c). The light curve of R2 has a similar intensity and profile as that of the southern ribbon as a whole, indicating that the emission from the straight segment dominates in the southern ribbon. However, the light curve of the hooked segment, despite being weaker, shows three peaks with increasing intensity at 06:57, 07:07, and 07:30 UT, respectively, with the late peak at 07:30 UT being dominant. Dividing the hooked segment into two with (R1a) and without (R1b) significant mass drainage, we found that it is the segment without significant mass drainage (R1b) that dominates the hooked segment (Fig.~\ref{fig:33}d). The light curve of the mass-draining segment (R1a) shows only one small peak at 07:07 UT, at the same time as the main peak of the total ribbon-swept area. Consistent with the 2011-11-09 event, the current event also suggests that the surface brightening caused by the impact of falling filament material is minor compared with that by magnetic reconnection. Hence, the different peak brightness observed in the two ribbon segments may result from different reconnection rates, but not from mass draining.

\section{Conclusion \& Discussion} \label{sec:discussion}

We studied three filament eruptions, each associated with a flare with two ribbons, whose opposite far ends are hooked, hence the 3D standard flare model can be applied to understand these events; i.e., the hooked ribbon maps the footprint of the QSL wrapping around the flux rope that embeds the filament, while the straight ribbon maps the footprint of the HFT consisting of two QSLs that intersects below the flux rope. For each eruption, we observed that most of the draining material flow downward in the form of vertical threads landing on the front of the dynamically evolving hooked ribbon. A small fraction of the falling material lands inside the region as half enclosed by the hooked ribbon, and the impact sites are sometimes manifested as discrete brightening points in EUV. In addition to flowing along filament legs, some material moves along curved trajectories and lands outside the two ribbons, with an additional ribbon forming at the impact site. On the other hand, we observed no falling material landing on the straight segments of the two ribbons. The post-flare arcade connecting the straight segments of flare ribbons is primarily produced by the reconnections that involve arcade field lines threading the vertical current sheet beneath the flux rope, and are termed `aa-rf' reconnection by \cite{Aulanier&Dudik2019}. Naturally, such field lines carry little dense filament material, and the absence of falling filament material on the straight segments of flare ribbons is consistent with the standard model. 

\begin{figure}
\plotone{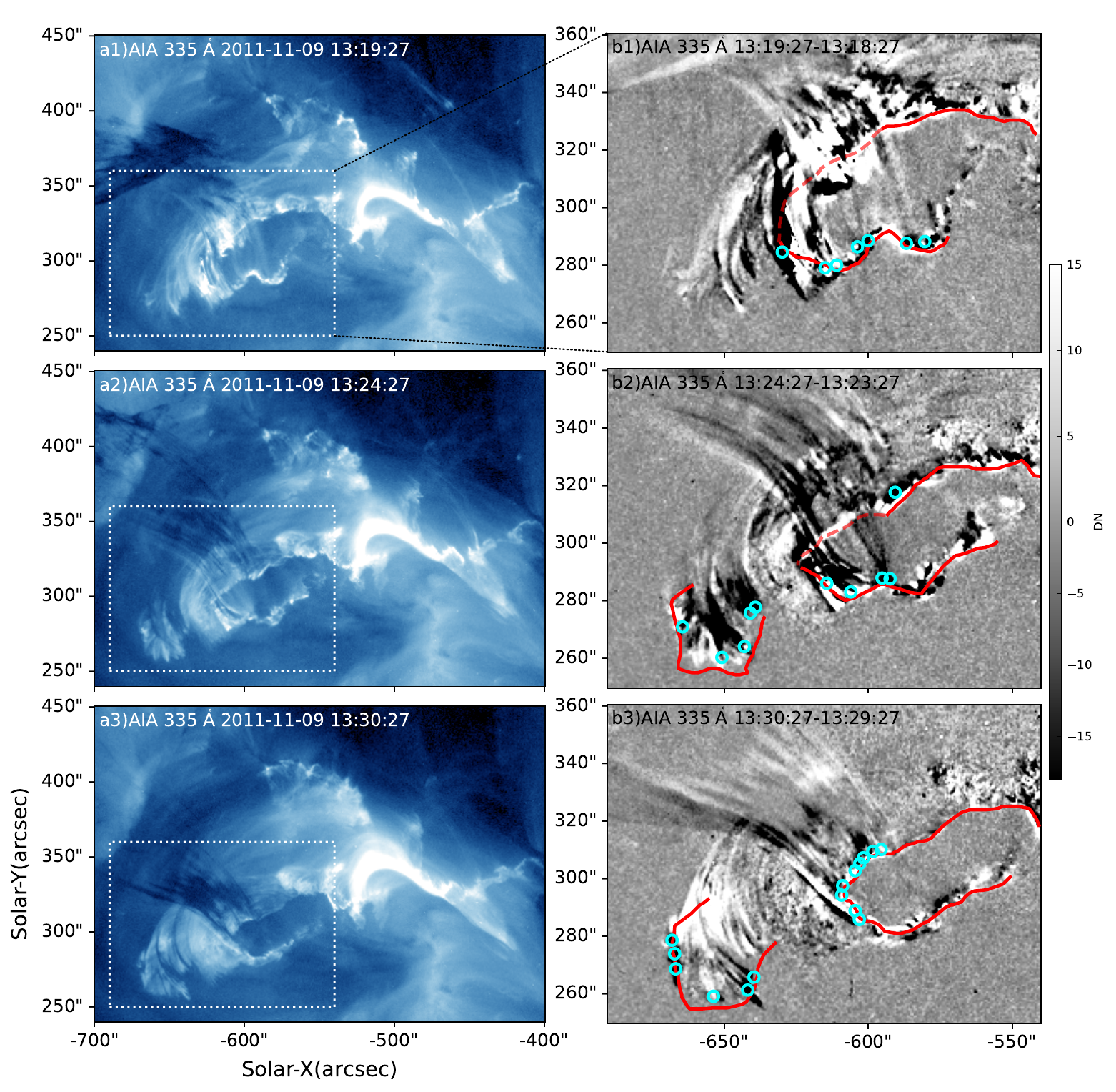}
\caption{Filament eruption on 2011 November 9. Three snapshots of the event in 335~{\AA} are shown in panels (a1--a3). The flare ribbons observed in 1600~{\AA} are outlined by red curves and superimposed on the 335~{\AA} difference images shown in (b1--b3), whose field-of-view is indicated by the white box in (a1--a3). The landing sites of falling threads are marked by cyan circles. The ribbon segments that are obscured by falling threads are indicated by dashed lines. The animation (from 12:51 UT to 13:59 UT) shows the multi-wavelength evolution of this event and has a duration of 10 s.\\
(An animation is available online.) \label{fig:11}}
\end{figure}

\begin{figure}
\plotone{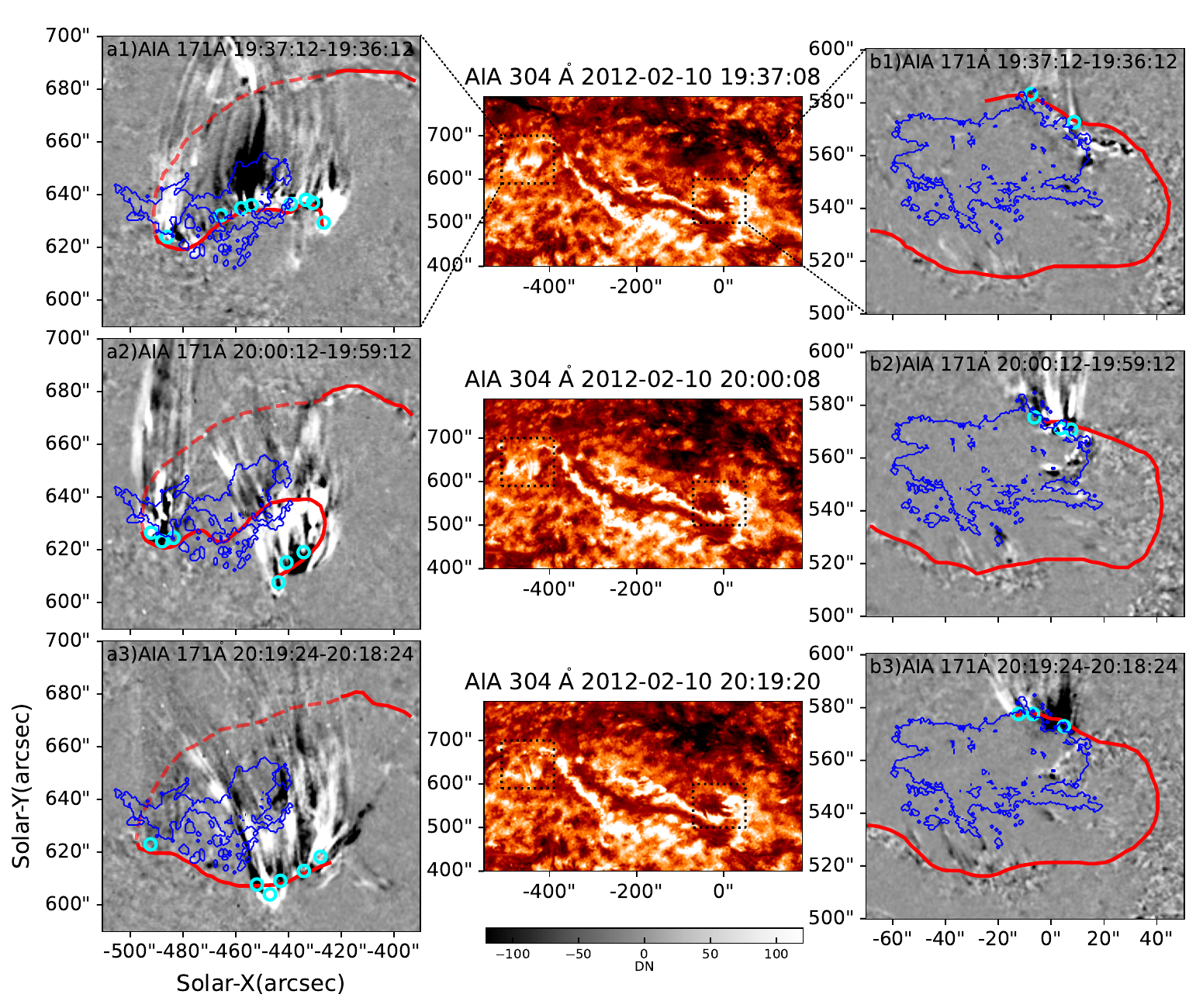}
\caption{Filament eruption on 2012 February 10. Three snapshots of the event in 304~{\AA} are shown in panels in the center column. The flare ribbons observed in 304~{\AA} are outlined by red curves and superimposed on the 171~{\AA} difference images shown in panels in the left and right columns. The landing sites of falling threads observed in 171~{\AA} are marked by cyan circles. The blue contours outline the coronal dimmings observed in 304~{\AA} to elucidate the two hooked regions. The animation (from 18:31 UT to 20:29 UT) shows the multi-wavelength evolution of this event and has a duration of 14 s.\\
(An animation is available online.) \label{fig:12}}
\end{figure}

\begin{figure}
\plotone{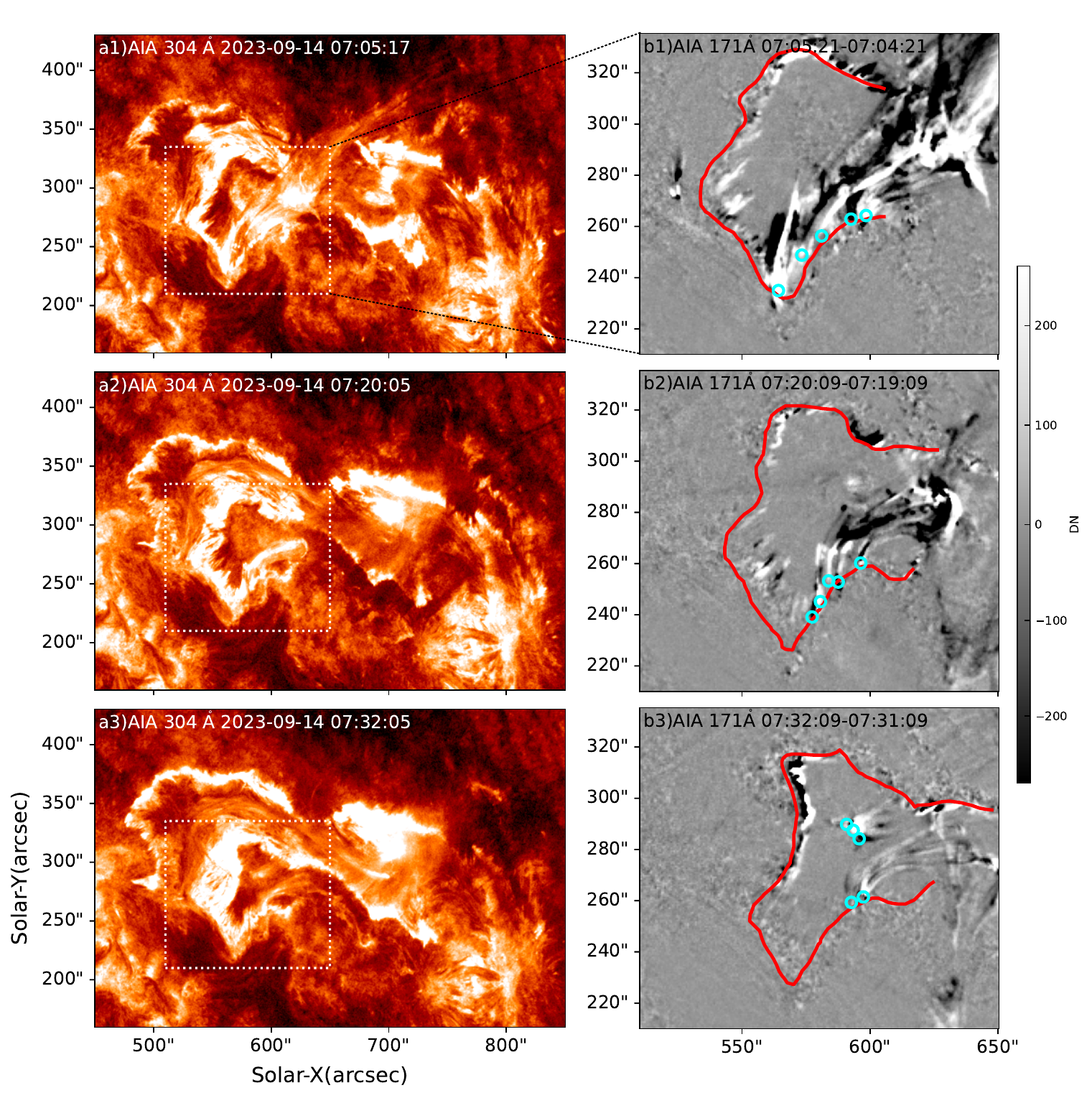}
\caption{Filament eruption on 2023 September 14. Three snapshots of the event in 304~{\AA} are shown in panels (a1--a3). The flare ribbons observed in 1600~{\AA} are outlined by red curves and superimposed on the 171~{\AA} difference images shown in (b1--b3), whose field-of-view is indicated by the white box in (a1--a3). The landing sites of falling threads observed in 171~{\AA} are marked by cyan circles.The animation (from 06:31 UT to 08:29 UT) shows the multi-wavelength evolution of this event and has a duration of 14 s.\\
(An animation is available online.)\label{fig:13}}
\end{figure}

\begin{figure}
\plotone{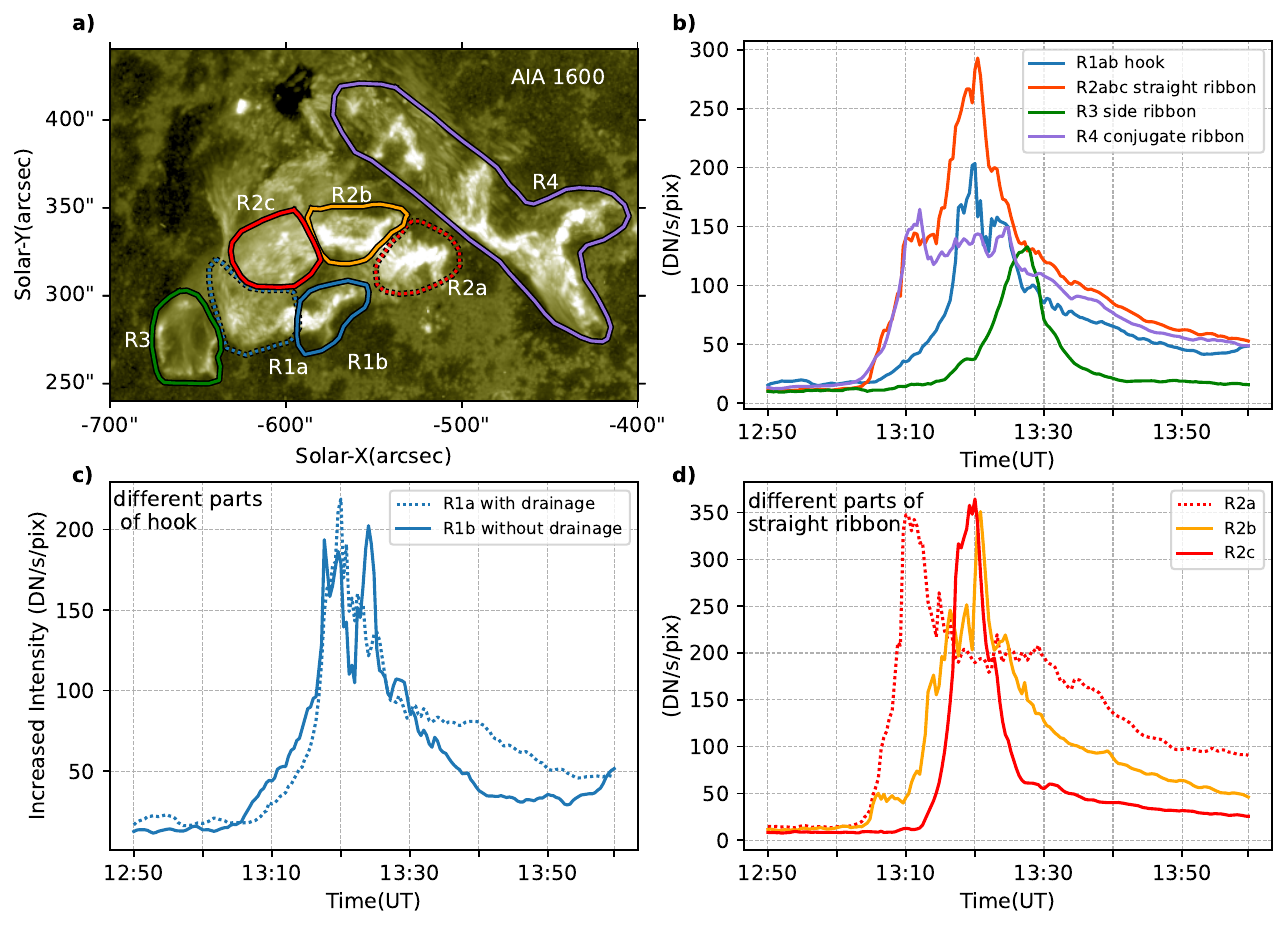}
\caption{Light curves of different segments of the flare ribbons in the 2011-11-09 event. The background intensity has been subtracted from the light curves.  (a) synoptic map of flare ribbons in 1600~{\AA}, with each pixel shown by its maximum intensity during the flaring period. (b) light curves for the hooked (R1) and straight (R2) ribbon in the south, side ribbon (R4), and the conjugated ribbon in the north (R4), respectively. (c) light curves for segments of the hooked ribbon with and without drainage. (d) light curves for different segments of the straight ribbon in the south as outlined in (a). 
\label{fig:31}} 
\end{figure}

During the eruption, filament materials would slide down along magnetic field lines under the force of gravity, whether they are situated in magnetic dips or not, because even dips would disappear when the field lines are stretched by the rising of the filament, or reconnected with the ambient fields. A flux rope rising fast enough may carry some filament mass into the higher corona or even the interplanetary space \cite[e.g.,][]{Sharma&Srivastava2012}. If, before eruption, a significant fraction of the filament mass were situated at the inner layers of the flux rope, we would have seen falling filament material land inside the hooked region. In contrast, most of the falling material lands right onto the hooked ribbon, with the impact sites coinciding with the ribbon fronts undergoing sweeping motions (see the animations accompanying Figures~\ref{fig:11}-\ref{fig:13}). Hence we conclude that 1) initially most of the filament mass is entrained by field lines threading the QSL boundary of the filament field, and that 2) magnetic reconnection that involves these field lines is the major cause of the mass drainage. The reconnection most likely occurs between a mass-carrying field line threading the flux-rope boundary and an arcade field line, consequently producing a new flux-rope field line and a flare loop, which is termed `ar-rf' reconnection by \cite{Aulanier&Dudik2019}. One expects that the filament material would fall not only onto the footpoints of the old flux-rope field line, which are located on the hooked ribbon, but also onto the footpoints of the arcade field line, thereby producing remote brightenings, such as the `side ribbon' in the 2011-11-09 event (Figure~\ref{fig:31}). In retrospect, the falling threads could be observed as ``spike- or fan-shaped brightenings that appear to mark the far endpoints of the filament'' \cite[]{Wang2009} in images with inferior resolution ($\sim5''$) and cadence (12 min). \cite{Wang2009} interpreted the brightenings as the ``the footprints of the current sheet formed at the leading edge of the erupting filament'', which is consistent with our interpretation based on the landing sites of falling threads. 

In the standard model \cite[]{Lin&Forbes2000}, flaring reconnections in the vertical current sheet add layers of flux wrapping around the original flux rope, which makes the rope's footpoints expand \cite[e.g.,][]{Wang2017}. In contrast, `ar-rf' reconnections `peel' flux away from the original flux rope by changing magnetic connectivities of the field lines threading its QSL boundary. Such reconnections erode the rope, making its footpoints shrink, which would be observed as the hooked ribbon sweeping inward. When some reconnections erode the rope from one side while others expand it from the other side, the rope's footpoints could drift continuously \cite[]{Aulanier&Dudik2019,Dudik2019} or even `jump' drastically \cite[]{Gou2023}.

On the hooked ribbon, we find that the ribbon segment that is not associated with significant mass drainage shows a delayed peak of the ribbon brightness in comparison to the segment associated with significant mass drainage. Since the hooked ribbon, which maps the QSL boundary of the mass-carrying flux rope, is undergoing inward sweeping motions in the 2011-11-09 and 2023-09-14 events (see the animation accompanying Figures~\ref{fig:11} and \ref{fig:13}), we infer that the `ar-rf' reconnection must be dominant as the hooked region shrinks. If we take the two peaks at the hooked ribbon as an indicator of two major episodes of `ar-rf' reconnection, both of which occur at the dynamically evolving QSL boundary, then the earlier (later) formed QSL boundary must be threaded by field lines loaded with (without) a significant amount of filament material. This is also consistent with our conjecture that initially most of the filament mass is embedded at the bottom boundary of the flux rope.    

Additionally, we modeled the constant-speed falling of filament material during eruption by assuming that the effective gravity of an obliquely falling cylinder is balanced by the drag force. Based on this simplified model, we are able to derive the density contrast of the falling filament threads relative to the background coronal plasma. Further, we carried out numerical experiments on the falling process by considering only gravity and pressure gradient forces, and found that the final speed is sensitive to the initial density contrast, which can be modulated by initial heights and thread density. The fact that the simulated velocity falls short of the observed ones suggests that we may have underestimated the initial height or density contrast, or, that the falling material has an initial speed owing to magnetic reconnection. To fully understand the dynamics of falling threads, one may need to determine the thread parameters with better observations and more reliable methods, to adopt a more realistic atmosphere model, and to include non-ideal effects such as viscosity and magnetic resistivity.

\begin{acknowledgments}
This work was supported by the Strategic Priority Program of the Chinese Academy of Sciences (XDB0560102), the National Key R\&D Program of China (2022YFF0503002), and the NSFC (42274204, 12373064, 42188101, 11925302).
\end{acknowledgments}

\bibliography{sample631}{}
\bibliographystyle{aasjournal}

\end{document}